\documentclass[prb,preprint]{revtex4-1} 

\usepackage{amssymb}
\usepackage{amsfonts}
\usepackage{amsmath,color}
\usepackage[dvips]{graphicx}
\usepackage{bm}

\def\mb#1{\mathbf{#1}}

\def\ber{\begin{eqnarray}}
\def\eer{\end{eqnarray}}
\def\beq{\begin{equation}}
\def\eeq{\end{equation}}

\def\ed{\end{document}}

\def\dT#1{\frac{\mathrm{d} #1}{\mathrm{d}T}}
\def\dTT#1{\frac{\mathrm{d} ^{2}#1}{\mathrm{d}T^{2}}}
\def\dt#1{\frac{\mathrm{d} #1}{\mathrm{d}t}}
\def\dtt#1{\frac{\mathrm{d} ^{2}#1}{\mathrm{d}t^{2}}}

\def\dtau#1{\frac{\mathrm{d} #1}{\mathrm{d}\tau}}
\def\dttau#1{\frac{\mathrm{d} ^{2}#1}{\mathrm{d}\tau^{2}}}
\def\sT{\sin \left(\omega T \right)}
\def\cT{\cos \left(\omega T \right)}

\begin{document}

\author{Matteo Luca Ruggiero}
\email{matteo.ruggiero@polito.it}
\affiliation{Politecnico di Torino, Corso Duca degli Abruzzi 24,  10129 Torino, Italy }
\date{\today}

\title{Gravitational waves physics using Fermi coordinates: a new teaching perspective}

\begin{abstract}
The detection of gravitational waves is possible  thanks to a multidisciplinary approach, involving different disciplines such as  astrophysics, physics, engineering and quantum optics. Consequently, it is important today for teachers to introduce the basic features of gravitational waves science
in the undergraduate curriculum. The usual approach to gravitational wave physics is based on the use of traceless and transverse coordinates, which do not have a direct physical meaning and, in a teaching perspective,  may cause misconceptions. In this paper, using Fermi coordinates, which  are simply related to observable quantities, we show that it is possible to introduce a gravitoelectromagnetic analogy that describes the action of  gravitational waves on test masses in terms of electric-like and magnetic-like forces.  We suggest that this approach could be more suitable when introducing the basic principles of gravitational waves physics to students.
\end{abstract}

\maketitle

\section{Introduction}\label{sec:intro}

In the framework of  General Relativity (GR) Einstein \cite{Einstein:1916cc,Einstein:1918btx} calculated the emission of gravitational waves (GWs) on the basis of the quadrupole formula. According to this formula, ideal candidates for the emission of GWs are huge masses moving at highly relativistic velocities. These sources are far away from the Earth, and their effects are detected as small perturbations or ripples in the fabric of space-time.  The first indirect evidence of the existence of GWs was obtained studying the binary pulsar B1913+16:\cite{Hulse:1974eb} the orbits of binary systems are modified  by the emission of gravitational waves, and these modifications can be observed by accurate timing measurements \cite{Weisberg:2004hi}.  It took more or less 100 years after Einstein's first calculation to obtain, in 2015, the first direct detection of GWs \cite{abbott2016observation}, which constituted the birth  of \textit{gravitational waves astronomy}.  We are currently in the era of multi-messenger astronomy: a given astrophysical source can be detected by means of different messengers.\cite{burns2019opportunities} Therefore, it is  important today for physics and astronomy teachers to explain the foundations of GWs science, starting from the emission process up to their detection, which requires a  multidisciplinary approach involving also engineering and quantum optics.

There are concrete proposals to integrate GWs science into physics and astronomy curricula.\cite{Farr:2011be} To this end, several introductory textbooks can be used, such as Refs.\ \onlinecite{d1992introducing,schutz2009first}; a comprehensive collection of useful materials  can be found in the resource letter in Ref. \ \onlinecite{Centrella:2003gs}.  There is also literature focusing on specific issues: for instance, Ref. \onlinecite{schutz1984gravitational}   suggests that the basic properties of GWs can be obtained by combining Newtonian gravity with the retardation effects due to the finite size of the speed of light. The detection process is made clear by considering the basic principles of interferometric detectors, as explained by Refs.\ \onlinecite{Saulson:1997cg,Melissinos:2010bv}, while Ref.\ \onlinecite{Mathur:2017cs}  discusses the key features of LIGO  on the basis of Newtonian mechanics, dimensional considerations, and analogies between gravitational and electromagnetic waves. The mechanism of emission can be studied on  the basis of the post-Newtonian theory, as discussed in Ref. \ \onlinecite{Buskirk:2019ho}, while data analysis can be used as a tool to design a GWs laboratory, as suggested in Ref. \ \onlinecite{Burko:2017ex}.

There are subtle issues connected with the process of detection of GWs, due to the distinct role of coordinates and observable quantities in GR.\cite{Garfinkle:2006bz} In Einstein's theory, physical measurements are meaningful only when the observer and the object of the observations are unambiguously identified.\cite{de2010classical} Roughly speaking, there are three steps in the measurement process: (i) observers possess their own space-time, in the vicinity of their world-lines; (ii) covariant physics laws are  projected onto local space and time; (iii) predictions for the outcome of measurements  in the local space-time of the observers are obtained.  Gravitational waves are usually described in terms of a transverse and traceless (TT)  tensor, which allows to introduce the so-called TT coordinates (see e.g. Ref.\ \onlinecite{Rakhmanov_2014} and references therein for a thorough discussion on the various coordinates used to describe the interaction with GWs). TT coordinates are used because they do not contain gauge-depending information; however, from a teaching perspective, TT coordinates  are difficult to handle, since they are not strictly related to measurable quantities and lack a direct physical meaning.\cite{flanagan2005basics}

When he was 21 (some months before obtaining his undergraduate degree in physics),  Enrico Fermi published an influential paper.\cite{1922RendL..31...21F} In this paper he introduced a quasi-Cartesian coordinates system in the observer's neighbourhood to describe the effects of gravitation. Such a set of coordinates, called \textit{Fermi coordinates}, adapted to the world-line of an observer, defines a \textit{Fermi frame}. Using this approach it is simple to emphasise that what an observer measures depends both on the background field where he is moving and, also, on his  motion. This is quite similar to what happens when we study  classical mechanics in non inertial frames: inertial forces appear, depending on the peculiar motion of the frame with respect to an inertial one. Fermi coordinates have a concrete meaning, since they are the coordinates an observer would naturally use to make space and time measurements in the vicinity of his world-line. 
Fermi  coordinates are defined, by construction, as scalar invariants.\cite{synge1960relativity} They are of the utmost importance to
understand the measurement process in GR, which is relevant in experimental tests of gravity; moreover, they provide a simple interpretation of the equivalence principle.

It is possible to show  that using Fermi coordinates the effects of a plane gravitational wave can be described by gravitoelectromagnetic fields:\cite{Ruggiero_2020} in other words, the wave field is equivalent to the combined action of a gravitoelectric and a gravitomagnetic fields that are transverse to the propagation direction and orthogonal to each other.  The analogy between electromagnetic fields and gravitational fields was already envisaged by Heaviside, on the basis of the similarity between Newton's law of gravitation and Coulomb's law of electrostatic force (see e.g. Ref. \ \onlinecite{McDonald:1997fd} and references therein). General Relativity naturally predicts the existence of a gravitomagnetic field, produced by mass currents, in analogy to what happens for the magnetic field, produced by charge currents. Generally, it is possible to describe gravitational effects on the basis of a gravitoelectrogmagnetic   analogy as discussed in Refs. \ \onlinecite{Ruggiero:2002hz,Mashhoon:2003ax}.

In this paper we introduce a gravitoelectrogmagnetic description of the field of a plane gravitational wave using Fermi coordinates, and we show that, in doing so, it is possible to understand the basic features of GWs. In particular, this approach describes the interaction with  detectors in terms of a Lorentz-like force equation. Moreover, we show that,  while existing detectors, such LIGO and VIRGO, or future ones, such as LISA, reveal the interaction of test masses with the gravitoelectric components of the wave, there are also gravitomagnetic interactions that could be used to detect the effect of GWs on moving masses  and spinning particles.\cite{biniortolan2017,Ruggiero_2020,Ruggiero_2020b}

We believe that this approach can be useful for teaching gravitational waves physics because 
it directly leads to measurable quantities, avoiding possible misunderstanding deriving from  the use of other types of coordinates.  Making an analogy with the more familiar concept of electromagnetic waves can also help students understand the new concept of gravitational waves.\cite{duit1991role,venville1996role}

The paper is organized as follows: in section \ref{sec:GW} we briefly review the classical approach to gravitational waves, while we discuss Fermi coordinates and the gravitoelectrogmagnetic approach in section \ref{sec:fermi}. On the basis of the gravitoelectrogmagnetic analogy, we describe in section \ref{sec:stgem} the interaction of GWs with detectors. Conclusions are eventually drawn in section \ref{sec:conc}.\\

We use the convention in which {Greek indices refer to space-time coordinates and assume the values $0,1,2,3$, while Latin indices refer to spatial coordinates and assume the values $1,2,3$, usually corresponding to the Cartesian coordinates $x,y,z$; the spacetime signature is $(-1,1,1,1)$.}

\section{Gravitational Waves in Transverse, Traceless Gauge}\label{sec:GW}

In this section we briefly recall the standard approach to the description of gravitational waves. To this end, we start from Einstein's equations 
\beq
G_{\mu\nu}=\frac{8\pi G}{c^{4}}T_{\mu\nu}, \label{eq:einstein0}
\eeq
and we suppose that the space-time metric $g_{\mu\nu}$ is in the form $\displaystyle g_{\mu\nu}=\eta_{\mu\nu}+h_{\mu\nu}$, where $|h_{\mu\nu}|\ll 1$ is a small perturbation of the Minkowski tensor $\eta_{\mu\nu}$ of flat space-time. Setting $\bar h_{\mu\nu}=h_{\mu\nu}-\frac 1 2 \eta_{\mu\nu}h$, with $h=h^{\mu}_{\mu}$, Einstein's field equations (\ref{eq:einstein0}) in the Lorentz gauge $\displaystyle \partial_{\mu} \bar h^{\mu\nu}=0$ (where $\displaystyle \partial_{\mu} = \frac{\partial}{\partial x^{\mu}}$) turn out to be
\beq
\square \bar h_{\mu\nu}=-\frac{16\pi G}{c^{4}}T_{\mu\nu}, \label{eq:einstein2}
\eeq
where $\square = \partial_{\mu}\partial^{\mu}=\nabla^{2}-\frac{1}{c^{2}}\frac{\partial}{\partial t^{2}}$ is the d'Alambert operator. Gravitational waves propagate through empty space and are solutions of equations (\ref{eq:einstein2}) in vacuum: 
\beq
\square \bar h_{\mu\nu}=0. \label{eq:einstein1}
\eeq
Typically, these equations are solved using  the so-called transverse - traceless coordinates    (see e.g. Ref. \ \onlinecite{Rakhmanov_2014}). In particular,   we look for plane wave solutions propagating  along the $x$ axis. Accordingly, a solution of (\ref{eq:einstein1}) can be written in the form
\beq
\bar h_{\mu\nu}=-\left(h^{+}e^{+}_{\mu\nu}+h^{\times}e^{\times}_{\mu\nu} \right), \label{eq:gwsol1}
\eeq
{with
\beq
h^{+}=A^{+}\cos \left(\omega t-kx +\phi^{+} \right), \quad h^{\times}=A^{\times}\cos \left(\omega t-kx +\phi^{\times}\right), \label{eq:gwsol20}
\eeq
where $\phi^{+}, \phi^{\times}$ are constants, and
\beq
e^{+}_{\mu\nu}=\left[\begin{array}{cccc}0 & 0 & 0 & 0 \\0 & 0 & 0 & 0 \\0 & 0 & 1 & 0 \\0 & 0 & 0 & -1\end{array}\right], \quad e^{\times}_{\mu\nu}=\left[\begin{array}{cccc}0 & 0 & 0 & 0 \\0 & 0 & 0 & 0 \\0 & 0 & 0 & 1 \\0 & 0 & 1 & 0\end{array}\right] \label{eq:gwsol3}
\eeq 
are the linear polarization tensors of the wave. In the above definitions, $A^{+}, A^{\times}$ are the amplitude of the wave in the two polarization states, $\phi^{+}, \phi^{\times}$ the corresponding phases,  while $\omega$ is the frequency and $k$ the wave number, so that the wave four-vector is $\displaystyle k^{\mu}=\left(\frac \omega c, k, 0, 0 \right)$, with $k^{\mu}k_{\mu}=0$. Notice that any radiation field of spin $s$ has two states of linear polarization inclined to each other at an angle of $\pi/(2s)$: for the photon $s=1$ and the linear polarization states of electromagnetic waves are orthogonal, while for gravitational waves  the linear polarization states (\ref{eq:gwsol3}) are at $\pi/4$, since $s=2$  (see  e.g. Ref. \ \onlinecite{MTW}). Furthermore, by analogy with electromagnetic waves, the two linear polarizations states can be added with phase difference of $\pm \pi/2$ to get circularly polarized waves. We will use 
\beq
h^{+}=A^{+}\sin \left(\omega t-kx  \right), \quad h^{\times}=A^{\times}\cos \left(\omega t-kx\right), \label{eq:gwsol2}
\eeq
thus fixing the phase difference: accordingly, circular polarization corresponds to the condition $A^{+}=\pm A_{\times}$. }

In TT coordinates the gravitational field of the wave is described by the line element
\beq
ds^2= -c^{2}dt^2+dx^2 +(1-h^{+})dy^2 +(1+h^{+})dz^2 -2h_{\times} dy dz\,. \label{eq:TTmetrica}
\eeq

Now, in order to understand the effect of a GW on test masses, we focus on the geodesic equation: starting from
\beq
\dttau x^{\mu}+\Gamma^{\mu}_{\alpha\beta} \dtau{x^{\alpha}} \dtau{x^{\beta}}=0,  \label{eq:geott1}
\eeq
and using the chain rule 
\beq
\dtau{x^{i}}=\dt{x^{i}}\dtau{t}\qquad   \dttau{x^{i}}=\dtt{x^{i}}\left(\dtau t \right)^{2}+\dt{x^{i}}\dttau t, \label{eq:geott10}
\eeq
the space components of the  equation (\ref{eq:geott1})  can be expressed in terms of the form:
\beq
\dtt{x^{i}} \left(\dtau t \right)^{2}+\dt{x^{i}}\dttau t +\Gamma^{i}_{\alpha\beta} \dt{x^{\alpha}}\dt{x^{\beta}} \left(\dtau t \right)^{2} =0 ,\label{eq:geott2}
\eeq 
where $\displaystyle a^{i}=\dtt{x^{i}}$ and  $\displaystyle v^{i}=\dt{x^{i}}$ are, respectively, the coordinate acceleration and velocity. If we also use the time component, we can write 
\beq
\frac{1}{c^{2}}\dtt{x^{i}}=-{\Gamma}^{i}_{jk}\frac{v^{j}}{c}\frac{v^{k}}{c}-2\Gamma^{i}_{0j}\frac{v^{j}}{c}-\Gamma^{i}_{00}+\frac{v^{i}}{c} \left(\Gamma^{0}_{00}+\Gamma^{0}_{jk}\frac{v^{j}}{c}\frac{v^{k}}{c}+2\Gamma^{0}_{0j}\frac{v^{j}}{c} \right). \label{eq:geott3}
\eeq
We suppose that test masses are moving at non-relativistic velocities,  which is reasonable if we are dealing with detectors: hence, since $\displaystyle \frac{|v^{i}|}{c} \ll 1$,  we can neglect all velocity-dependent terms in Eq. (\ref{eq:geott3}) and obtain
\beq
\frac{1}{c^{2}}\dtt{x^{i}}=-\Gamma^{i}_{00}.  \label{eq:geott4}
\eeq
Using the metric  (\ref{eq:TTmetrica}), we get   $\Gamma^{i}_{00}=0$ and, then,  $\displaystyle \dtt{x^{i}}=0$. {It is interesting to point out that this result is true for any gravitational field in the form
\beq
ds^{2}=-c^{2}dt^{2}+g_{ij}(x^{\mu})dx^{i}dx^{j}, \label{eq:metricageod}
\eeq
i.e. with $g_{00}=-1$ and $g_{0i}=0$: all test particles that are spatially at rest in such a spacetime follow geodesics. In TT gauge the metric (\ref{eq:TTmetrica}) is in the form (\ref{eq:metricageod}), and this  simply means that the TT coordinates of a test mass acted upon by a gravitational wave do not change}: but remember that coordinates in general relativity do not have a direct physical meaning. In order to see the effect of GWs on test masses, we need to evaluate the variation of the physical distance between them, which is defined by the  proper length and not by the coordinate distance. For instance,  let us suppose that two test masses are located along the $y$ axis, at $P_{1}=(0,0,0)$ and $P_{2}=(0,L,0)$. The proper distance $d_{y}$ between them is obtained from the line element (\ref{eq:TTmetrica}):
\beq
d_{y}=\int_{0}^{L} \sqrt{g_{yy}}dy = \int_{0}^{L} \sqrt{1-h^{+}}dy \simeq \left(1-\frac{h^{+}}{2}\right) L, \label{eq:deltaL}
\eeq
where we have taken into account  the smallness of the perturbation; hence, according to  (\ref{eq:gwsol2}) the proper distance changes with time, due to the passage of the GWs. Interferometers like LIGO and VIRGO are designed to measure this change. As we are going to show below, Fermi coordinates allow  a direct description of the effect of GWs on test masses, in terms of measurable quantities.

TT gauge is used because of its convenience:\cite{flanagan2005basics} as we said, in linearized theory it fixes all gauge freedom  so that the metric perturbations are physical and do not contain gauge-depending information; moreover, in this gauge it is manifest that the GWs have two polarization components and that they are transverse to the propagation direction.

For future convenience, we remember that, in linear approximation in the perturbation $h_{\mu\nu}$,  the Riemann curvature tensor is written in the form:
{
\beq
R_{\alpha\mu\beta\nu}=\frac 1 2 \left( h_{\alpha\nu,\mu\beta}+h_{\mu\beta,\nu\alpha}-h_{\mu\nu,\alpha\beta}-h_{\alpha\beta,\mu\nu} \right). \label{eq:riemann01}
\eeq
In particular, since in the TT metric $h_{i0}=0$ and $h_{00}=-1$, we obtain the following expression of the Riemann tensor 
\beq
R_{0\mu\beta\nu}=\frac 1 2 \left(h_{\mu\beta,\nu 0}-h_{\mu\nu,0\beta} \right) \label{eq:riemann03}
\eeq
which  will be used below.}


\section{Fermi Coordinates}\label{sec:fermi}

The space-time metric in Fermi coordinates, in the vicinity of a given observer's world-line, depends both on \textit{where} and \textit{how} the observer is moving. In other words, the background space-time and the type of motion within it determine the local  metric, whose general expression can be found in Ref.\ \onlinecite{Ruggiero_2020}. Here, since we are concerned with GWs effects, for the sake of simplicity we consider an observer freely falling in the field of  a plane gravitational wave: hence, Fermi coordinates are a geodesic coordinate system based on non rotating frame along  the observer's world-line (the \textit{reference} world-line).\cite{Mashhoon:1996wa,marzlin}  If we set Fermi coordinates $X^{\alpha} = \left(cT,X,Y,Z \right) = \left( cT,\mb X \right)$,  the metric can be expressed in a power series in $\mb X$ from the reference world-line, in the form:\cite{manasse1963fermi,MTW}
\beq
ds^{2}=-\left(1+R_{0i0j}X^iX^j \right)c^{2}dT^{2}-\frac 4 3 R_{0jik}X^jX^k cdT dX^{i}+\left(\delta_{ij}-\frac{1}{3}R_{ikjl}X^kX^l \right)dX^{i}dX^{j}. \label{eq:mmmetric}
\eeq
The above expression is valid up to quadratic displacements $|X^{i}|$ from the reference world-line. Notice that $R_{\alpha \beta \gamma \delta}=R_{\alpha \beta \gamma \delta}(T)$ is the Riemann curvature tensor evaluated along the reference geodesic, where $\mb X=0$, and it depends on $T$ only, which is the observer's proper time.

As discussed in Ref.\  \onlinecite{Mashhoon:2003ax}, neglecting the terms $g_{ij}$ related to the spatial curvature,  the space-time element (\ref{eq:mmmetric})  can be recast  in terms of the gravito-electromagntic  potentials $(\Phi, \mb A)$  
\begin{equation} 
ds^2=-\left(1-2\frac{\Phi}{c^2}\right)c^{2}dT^2-\frac{4}{c}({\mb 
A}\cdot d{\mb
X})dt+\delta_{ij}dX^idX^j \, , \label{eq:mmetric2}
\end{equation}
where  the   \textit{gravitoelectric}  potential $\Phi=\Phi (T, {\mb X})$ is
\beq
\Phi (T, {\mb X})=-\frac{c^{2}}{2}R_{0i0j}(T )X^iX^j, \label{eq:defPhiG}
\eeq
and the components of the \textit{gravitomagnetic}  potential  $\mb A=\mb A (T ,{\mb X})$ turn out to be
\beq
A^{}_{i}(T ,{\mb X})=\frac{c^{2}}{3}R_{0jik}(T )X^jX^k. \label{eq:defAG}
\eeq
In close analogy to electromagnetism, the gravitoelectric and gravitomagnetic fields $\mb E$ and  $\mb B$ are defined in terms of the potentials by
\begin{equation} {\mb E}=-\nabla \Phi 
-\frac{1}{c}\frac{\partial}{\partial T}\left( \frac{1}{2}{\mb
A}\right),
\quad {\mb B}=\nabla \times {\mb A}. \label{eq:defEB1}
\end{equation}
Using the definitions (\ref{eq:defPhiG}) and (\ref{eq:defAG}) we obtain (up to linear order in $|X^{i}|$) the following components
{
\beq
E^{}_i(T ,{\mb X})=c^{2}R_{0i0j}(T) X^j, \label{eq:defEIEG}
\eeq
and 
\beq
B^{}_i(T ,{\mb R})=-\frac{c^{2}}{2}\epsilon_{ijk}R^{jk}_{\;\;\;\; 0l}(T )X^l. \label{eq:defBIBG}
\eeq}

The electromagnetic analogy is  useful to describe the motion of free test particles: namely, the geodesic  equation of the space-time metric (\ref{eq:mmetric2})   can be written in the form of a Lorentz-like force equation\cite{Mashhoon:2003ax}
\beq
m\dTT{\mb X}=q_E{\mb E}+q_B\frac{\mb V}{c}\times {\mb B},  \label{eq:llorenz1}
\eeq
up to linear order in the particle velocity $ \mb V=\dT{\mb X}$, which is actually a \textit{relative velocity}.

In the Lorentz-like force equation, $q_{E}=-m$ is the gravitoelectric charge, and $q_{B}=-2m$ is the gravitomagnetic  one (the minus sign takes into account the fact that the gravitational force is always attractive). We notice that the ratio $\displaystyle  \frac{q_{B}}{q_{E}}=2$, since linearized gravity is a spin-2 field.

 As a consequence, the Lorentz-like force equation becomes
\beq
m\dTT{\mb X}=-m{\mb
E}-2m\frac{\mb V}{c}\times {\mb B}. \label{eq:lorentz}
\eeq

This equation is the key point of this paper; let us briefly comment on its meaning and implication on the study of GWs. According to our approach, we may say that, in  his reference frame, the observer studies the evolution of a test mass using Eq. (\ref{eq:lorentz}); in other words, the latter equation describes how the mass coordinates $X,Y,Z$ (which, by construction, measure proper distances away from the reference world-line) change due to the action of the gravitational field. 


It is important to emphasise that the effects expressed by the gravitoelectrogmagnetic fields $\mb E$ and $\mb B$ have a tidal character, since both fields (\ref{eq:defEIEG})-(\ref{eq:defBIBG}) depend on the location of the mass, \textit{relative} to the observer which is at the origin of the frame. Accordingly, the action of the GWs is  simply described in the Fermi frame in terms of Newtonian gravitoelectrogmagnetic forces; of course, if other forces are present (such as mechanical or electromagnetic ones) they should be added to the  equation of motion. 

The gravitoelectrogmagnetic fields $\mb E$ and $\mb B$, according to their expressions (\ref{eq:defEIEG})-(\ref{eq:defBIBG}), vanish along the reference world-line, where $\mb X=0$; then, the Lorentz-like force equation (\ref{eq:lorentz}) suggests  that test masses are freely moving. This is nothing but a rephrasing of the equivalence principle: in local {freely falling frames} the physics of special relativity holds true.

In the following section we are going to show how, using the Lorentz-like equation (\ref{eq:lorentz}) it is possible to describe the interaction of the wave with a detector. To be more specific, in the Riemann curvature tensor needed to the define the gravitoelectrogmagnetic fields, we will neglect the contributions  due to local gravitational fields (such as the one of the Earth) and we will consider only the contribution of the wave.

\section{Gravitoelectrogmagnetic effects in the Fermi frame} \label{sec:stgem}

According to what we have seen before, in order to study the interaction of GWs with test masses, we can use the {Lorentz-like} force equation (\ref{eq:lorentz}): to this end, we need the explicit expressions of the gravitoelectrogmagnetic fields appearing therein. We remember that these fields are  defined in terms of the Riemann tensor:
\beq
E^{C}_i(T ,{\mb X})=c^{2}R_{0i0j}(T) X^j, \quad B^{C}_i(T ,{\mb R})=-\frac{c^{2}}{2}\epsilon_{ijk}R^{jk}_{\;\;\;\; 0l}(T )X^l. \label{eq:gemfields}
\eeq

To calculate these fields, in principle we need the expression of the Riemann tensor in Fermi coordinates. However, in the weak field approximation -  that is to say up to linear order in $h_{\mu\nu}$ -  the Riemann tensor is invariant with respect to  coordinate transformations, hence it  has the same expression in terms of the new coordinates. As a consequence, we can use the TT values for the perturbations $h_{\mu\nu}$ given by Eq.  (\ref{eq:TTmetrica}) and express them in Fermi coordinates. Also, dealing with GWs, in what follows we suppose that the extension of the reference frame is much smaller than the wavelength, so that we may neglect the spatial variation of the wave field:  consequently the components of the Riemann tensor are evaluated at the origin of our frame, where $\mb X=0$. {If this condition is not fulfilled, it is necessary to use the expression of the Fermi coordinates valid at higher order in the distance from the reference world-line (see e.g. Ref. \ \onlinecite{marzlin}) and, as a consequence, additional terms will be present.}

Using  Eqs. (\ref{eq:TTmetrica}) and (\ref{eq:riemann03}), the components of the gravitoelectric field (\ref{eq:defEIEG})  are
\small
\beq
E^{}_{X}  = 0, \quad E^{}_{Y}  = -\frac{\omega^{2}}{2}\left[A^{+} \sin \left(\omega T \right)Y+A^{\times} \cos \left(\omega T \right) Z \right], \quad E^{}_{Z}  = -\frac{\omega^{2}}{2}\left[A^{\times}\cT Y-A^{+}\sT Z \right], \label{eq:campoE}
\eeq
\normalsize
while those of the gravitomagnetic field (\ref{eq:defBIBG}) turn out to be
\small
\beq
B^{}_{X}  = 0, \quad B^{}_{Y}  = -\frac{\omega^{2}}{2}\left[-A^{\times} \cT Y+A^{+} \sT Z \right], \quad B^{}_{Z}  = -\frac{\omega^{2}}{2}\left[A^{+}\sT Y+A^{\times}\cT Z \right]. \label{eq:campoB}
\eeq
\normalsize

Notice that both fields are perpendicular to the propagation direction: gravitational waves, like electromagnetic ones, are transverse.
{Taking into account the expressions (\ref{eq:campoE}) and (\ref{eq:campoB}) it is easy to check that $\mb E \cdot \mb B =0$: in other words the two fields are orthogonal everywhere; moreover, we obtain also that $|\mb E|^{2}-|\mb B|^{2}=0$: they have the same magnitude.} Notice also that $\mb E (A^{\times})=\mb B (A^{+})$ and $\mb E (A^{+})=-\mb B (A^{\times})$.

\begin{figure}[h]
\begin{center}
\includegraphics[scale=1.20]{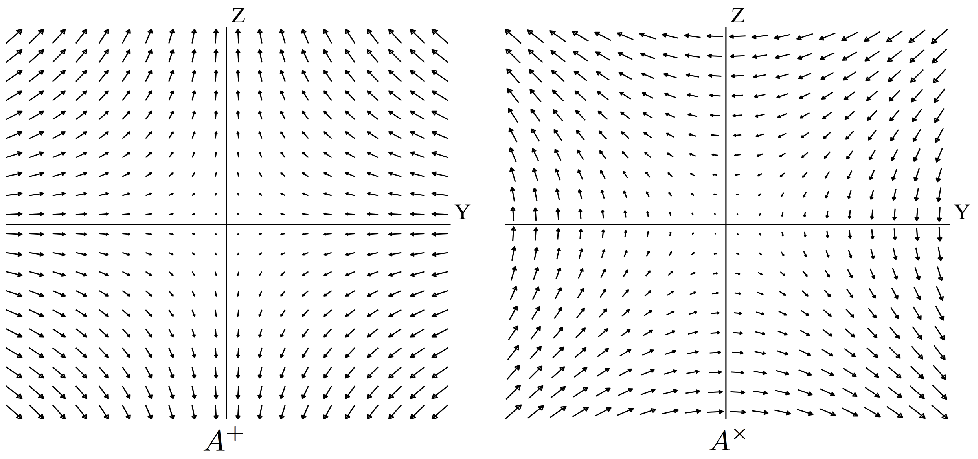}
\caption{The gravitoelectric field $\mb E$: on the left,  we suppose that the GW has just the $A^{+}$ polarization; on the right, we suppose that the GW has just the $A^{\times}$ polarization. Notice that the two polarizations differ by a rotation of $\pi/4$.} \label{fig:campoE}
\end{center}
\end{figure}

\begin{figure}[h]
\begin{center}
\includegraphics[scale=.60]{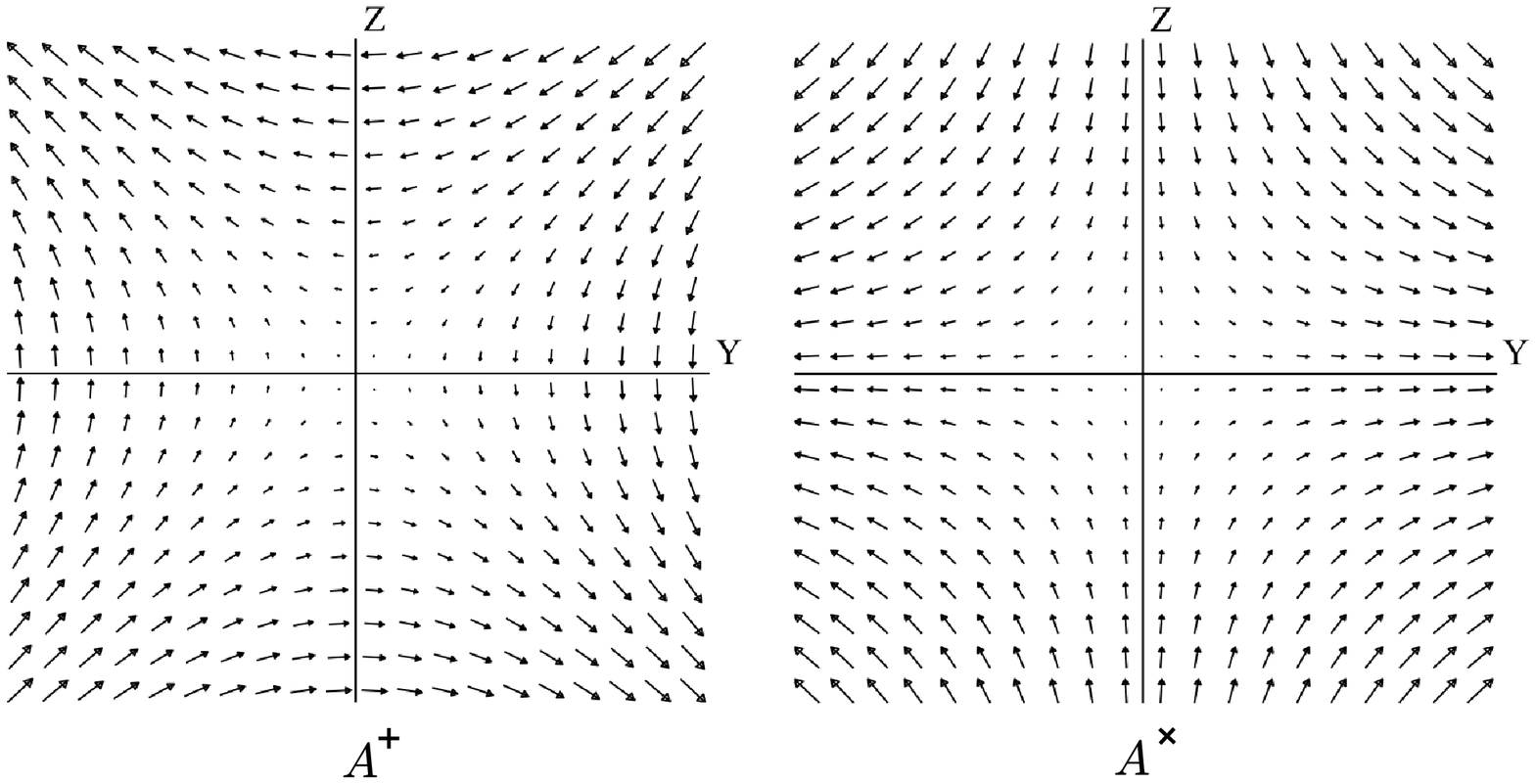}
\caption{The gravitomagnetic field $\mb B$: on the left,  we suppose that the GW has just the $A^{+}$ polarization; on the right, we suppose that the GW has just the $A^{\times}$ polarization. Notice that the two polarizations differ by a rotation of $\pi/4$.} \label{fig:campoB}
\end{center}
\end{figure}

\begin{figure}[h]
\begin{center}
\includegraphics[scale=.60]{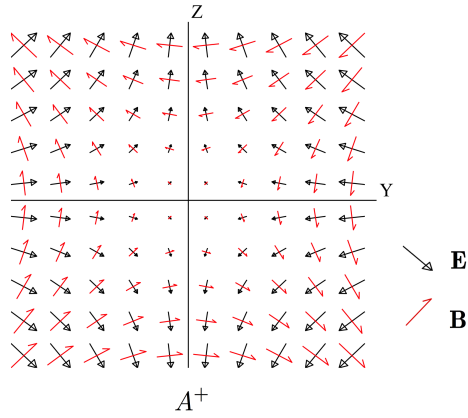}
\caption{{The gravitoelectric $\mb E $ and gravitomagnetic field $\mb B$ for a wave with $A^{+}$ polarization: notice that  the two fields  orthogonal  everywhere.} } \label{fig:campoEB}
\end{center}
\end{figure}

In Figures \ref{fig:campoE} and \ref{fig:campoB} the components of the gravitoelectric and gravitomagnetic fields are plotted at fixed $T$: it is manifest that, for both fields, the $A^{\times}$ components are obtained from the $A^{+}$ with a rotation of $\pi/4$. In Figure \ref{fig:campoEB} we see that, at fixed time,  the $A^{+}$ components of the fields are orthogonal at any spatial location; the same is true for the  $A^{\times}$ components.

Now, using the gravitoelectrogmagnetic approach,  we want to study the interaction of the wave with a detector. To begin with, it is important to understand, in this context, the meaning of \textit{detector}. One of the first extensive analyses of gravitational waves detectors can be found in the paper by Press and Thorne;\cite{press1972gravitational} here a GW is  described as ``field of (relative) gravitational forces propagating with the speed of light''. 
This definition nicely fits our approach, where it gets a true operational meaning: in fact using Fermi coordinates relative to a given observer, physical quantities, such as displacements, are \textit{relative} to the reference world-line. In particular the gravitoelectric and gravitomagnetic fields are position dependent and act differently on test masses located at different positions, thus producing tidal effects. In summary, the passage of  GWs provokes a space-time deformation which can be described in terms of tidal forces due to the gravitoelectrogmagnetic fields. 

That being said, a detector or \textit{gravitational antenna} is a physical system made of test masses, on which the wave acts producing displacements and motion relative to the reference world-line.  Starting from the explicit expressions of the gravitoelectrogmagnetic fields (\ref{eq:campoE}) and (\ref{eq:campoB}), we are now in a position to use the Lorentz-like force equation (\ref{eq:lorentz}) to describe the effect of GWs on gravitational antennas. Before doing that, it is important to point out the limits of our approximation: we work at first order in the wave amplitude, so we have to deal with  equations in a self-consistent way.  If we suppose that $\mb V^{0}$ is the velocity of a test mass before the passage of the wave, the wave produces a change $\mb V(T)=\mb V^{0}+\delta \mb V (T) $, where the variation $\delta \mb V (T)$ is of the order of the wave amplitude $A$: $\delta \mb V (T) = O(A)$.  As a consequence, in the linear approximation  in the equation of motion (\ref{eq:lorentz}) we can neglect the contribution of the gravitomagnetic field if the test masses are at rest before the passage of the wave; things are different  if we consider masses in motion before the passage of the wave.

The simplest GW-antenna is made of two free masses that are at rest before the passage of the wave; in particular, we suppose that one of them is at the origin, so we are interested in the motion of the other mass. Accordingly,  the test mass is acted upon by the gravitoelectric field only, and its equation of motion is
\beq
\dTT{\mb X}=-{\mb E} \label{eq:motomassa}
\eeq
Let us suppose that the polarization of the wave is such that $A^{\times}=0$ (as we have seen before, the effect of the $A^{\times}$ polarization is qualitatively the same); according to Eq. (\ref{eq:campoE}) the gravitoelectric field is given by
\beq
E^{}_{X}  = 0, \quad E^{}_{Y}  = -\frac{\omega^{2}}{2}\left[A^{+} \sin \left(\omega T \right)Y \right], \quad E^{}_{Z}  = \frac{\omega^{2}}{2}\left[A^{+}\sT Z \right]. \label{eq:defExyzcircular}
\eeq

Let the location of the mass before the passage of the wave  be $\mb X_{0}=(0,L,0)$, so that the physical distance between the two masses is $L$. Then, the the solution of Eq.  (\ref{eq:motomassa}) up to linear order in the wave amplitude, is
\beq
X(T)=0, \quad Y(T)=L\left[1-\frac{A^{+}}{2} \sT \right], \quad Z(T)=0. \label{eq:solmotomassa}
\eeq
The distance between the two masses changes with time. This result is in agreement with
Eq. (\ref{eq:deltaL}), obtained using TT coordinates and considering, then, the physical distance:   the coordinate distance in Fermi coordinates is an observable quantity.

Another kind of detector, called a heterodyne antenna,  was proposed during the 70's by Braginskij and collaborators; \cite{braginskij1969reception,braginsky1971heterodyne}  its functioning is based on the resonance principle. It is interesting to see how our approach allows to simply understand the interaction of  GWs with this device, using the equations of basic mechanics in the Fermi frame. 

\begin{figure}[h]
\begin{center}
\includegraphics[scale=0.60,draft=false]{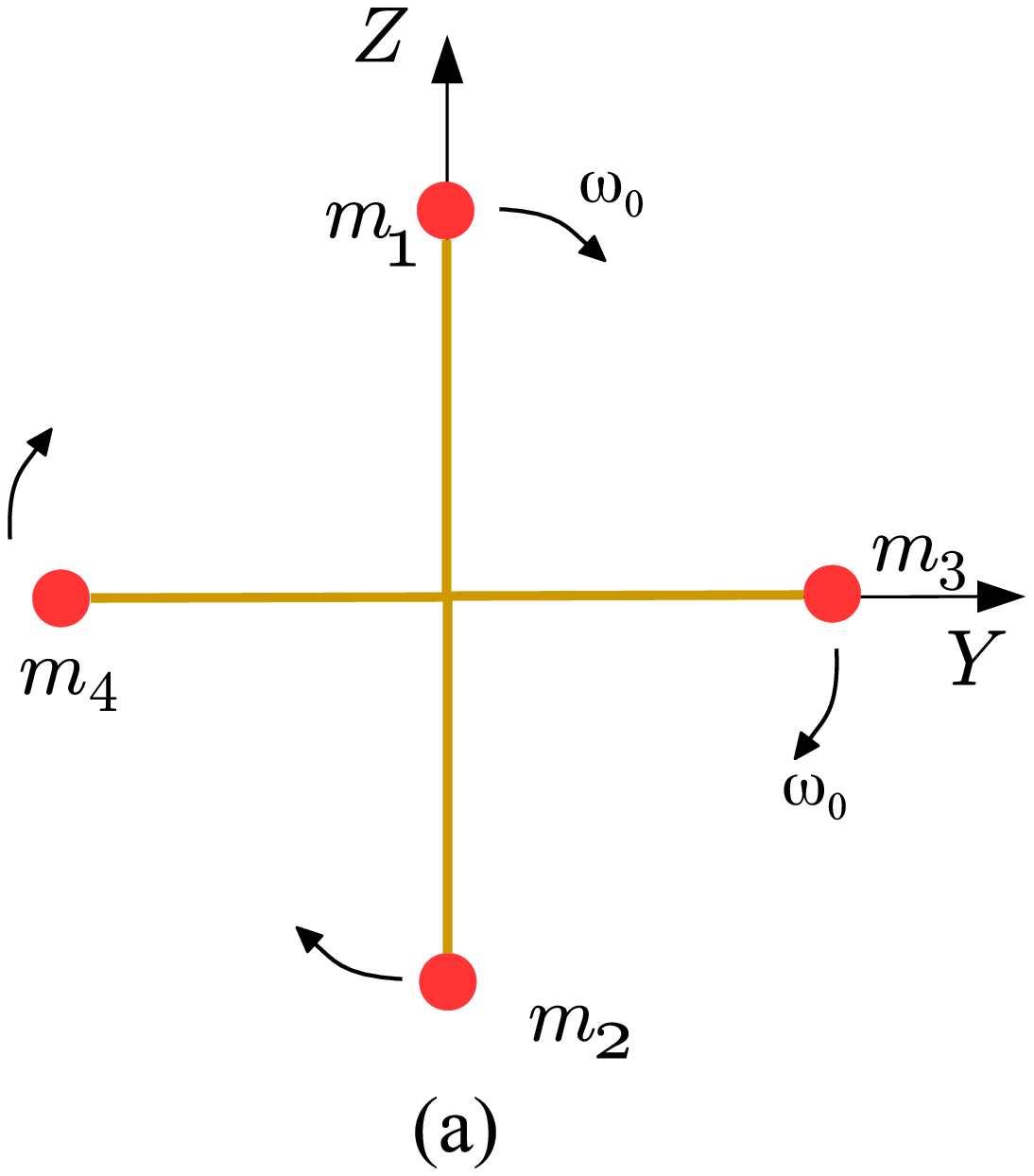}
\includegraphics[scale=0.60,draft=false]{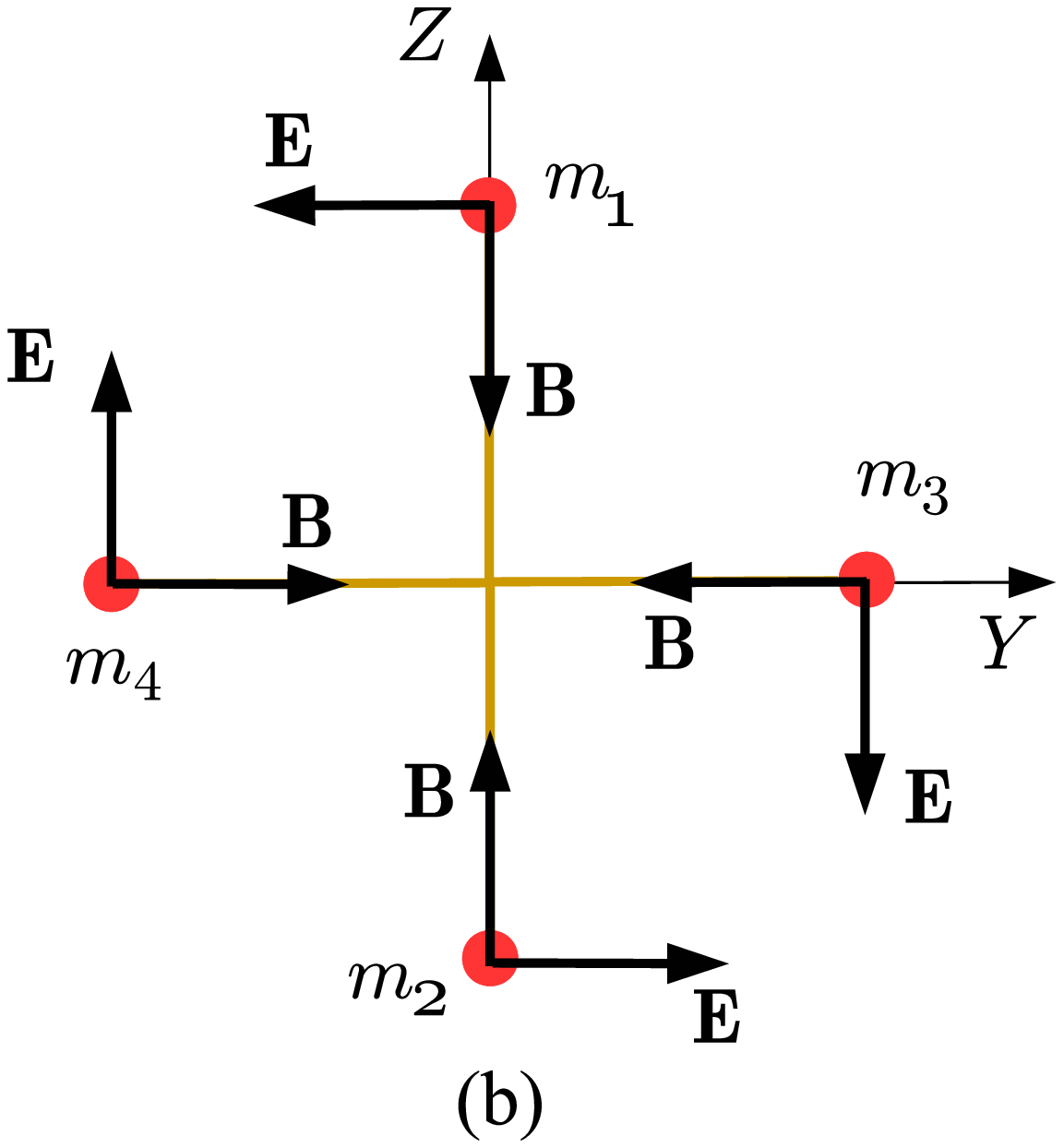}
\caption{{(a): two identical dumbbells are made by test masses $m_{1}=m_{2}=m_{3}=m_{4}=m$ at fixed distance $R$, and rotated by $\pi/2$ with respect each other. They independently rotate with frequency $\omega_{0}$, before the passage of the wave. (b): the gravitoelectromagnetic fields acting on the test masses, due the passage of the wave.} } \label{fig:dumb}
\end{center}
\end{figure}

The antenna is made of two dumbbells crossed at an angle of $\pi/2$, with length $R$. Before the passage of the wave, they independently rotate in the plane orthogonal to the propagation direction with the same frequency $\omega_{0}$.  We suppose that at $T=0$ the configuration of the dumbbells is that of Figure   \ref{fig:dumb}-(a), i.e. the four masses $m_{1}=m_{2}=m_{3}=m_{4}=m$ are along the axes $Y$ and $Z$. The coordinates of the mass $m_{1}$,  whose position at $T=0$ is $\mb X_{0}=(0,0,R)$, are 
\beq
X_{1}=0, \quad Y_{1}=R \sin \omega_{0}T, \quad Z_{1}=R \cos \omega_{0}T. \label{eq:dumb1}
\eeq
We suppose that the wave is circularly polarised, so that $A^{+}=A^{\times}=A$; using the gravitoelectric field (\ref{eq:campoE}),  the 
force acting on the mass is $\mb F^{E}_{1}=-m\mb E$ 
\beq
 F^{E}_{1,X}=0, \quad F^{E}_{1,Y}= \frac{m\omega^{2}A R}{2} \cos \left(\omega-\omega_{0} \right)T, \quad {F}^{E}_{1,Z}=- \frac{m\omega^{2}A R}{2} \sin \left(\omega-\omega_{0} \right)T. \label{eq:dumb2}
\eeq
If $\omega_{0}=\omega/2$, the above expression becomes
\beq
F^{E}_{1,X}=0, \quad F^{E}_{1,Y}= \frac{m\omega^{2}A R}{2} \cos\frac{\omega}{2} T, \quad F^{E}_{1,Z}=- \frac{m\omega^{2}A R}{2} \sin \frac{\omega}{2} T \label{eq:dumb3}
\eeq
Hence, if the \textit{resonance condition} $\omega_{0}=\omega/2$ is fulfilled,  the mass $m_{1}$ experiences a  force of  constant magnitude $|\mb F^{E}_{1}|=\frac{m\omega^{2}A R}{2}$, orthogonal to the dumbbell. A similar approach suggests that the other mass $m_{2}$  undergoes an equal force in opposite direction. {In summary, due to the action of the wave, a constant torque $\bm \tau_{12} = -m\omega^{2}A R^{2} \mb u_{X}$ (where $ \mb u_{X}$ is the unit vector of the $X$ axis) acts on the dumbbell, with the effect of \textit{accelerating} its rotation.} 

Applying the same approach to the other dumbbell, we see that it is acted upon by a constant torque  $\bm \tau_{34} = m\omega^{2}A R^{2} \mb u_{X}$, with the effect of \textit{decelerating} its rotation.

In summary, with this choice of the rotation frequency,  one dumbbell is accelerated and the other is decelerated, so that the masses come closer: the angular separation $\theta$ between the two dumbbells evolves with time with the law $\Delta(\theta)(T)=\frac{\pi}{2}-\delta \theta (T)=\frac{\pi}{2}-\frac 1 2 \omega^{2}A T^{2} $, which is independent of the length $R$.

However, our approach  based on the Lorentz force equation (\ref{eq:lorentz})  suggests that, since before the passage of the wave the masses are in motion, there is an additional effect, due to the action of the gravitomagnetic field on the rotating masses. The  gravitomagnetic force acting on a mass moving with speed $\mb V$, is $\mb F^{B}=-2m\frac{\mb V}{c}\times {\mb B}$. Since we are working at linear order in the wave amplitude, we use in this expression the velocity of the system before the passage of the wave. Let the rotation frequency be $\omega/2$ again:  from Eq. (\ref{eq:campoB}), we obtain  the gravitomagnetic field acting on the  the mass $m_{1}$:
\beq
B_{X}=0, \quad B_{Y}=-\frac{\omega^{2}AR}{2} \sin \frac{\omega }{2}T, \quad B_{Z}=-\frac{\omega^{2}AR}{2} \cos \frac{\omega }{2}T. \label{eq:dumb4}
\eeq
This field has constant magnitude and it is  always directed toward the center; hence, the mass $m_{1}$ undergoes the force $\mb F_{1}^{B}=\frac{m\omega^{3}AR^{2}}{2c} \mb u_{X}$. The other mass $m_{2}$ undergoes to the same force, so that the total force acting on the first dumbbell is $\mb F^{B}_{12}=\frac{m\omega^{3}AR^{2}}{c}\mb u_{X}$. If we consider the other dumbbell, using the same approach we see that it {experiences} a total force $\mb F^{B}_{34}=-\frac{m\omega^{3}AR^{2}}{c}\mb u_{X}$. The first dumbbell moves in the direction of propagation of the wave, while the other one moves in the opposite direction: accordingly,  their distance $d$ changes with time according to $d(T)=\frac{ \omega^{3}AR^{2}}{c} T^{2}$.   

{This effect  can be understood taking into account the expression of the Poynting vector
\beq
{\mb P}= \frac{c}{4\pi G} \mb E \times \mb B. \label{eq:poynting1}
\eeq
which defines the energy per unit time and unit of surface transported by the wave along its  propagation direction (see e.g. Ref.\ \onlinecite{Mashhoon:2003ax}).
In fact, if we consider the fields acting on the test masses, as described in Figure \ref{fig:dumb}-(b), it is easy to check that the Poynting vector acting on the first dumbbell is directed along the direction of propagation of the wave, while it acts in the opposite direction on the second dumbbell. Since Fermi coordinates have a concrete meaning and are strictly related to measurable quantities, this result is a simple demonstration of the wave transmitting linear momentum.    The effect of the wave on a rotating detector suggests another simple argument that proves the reality of gravitational waves in addition to famous \textit{sticky bead argument}, developed by Feynman and Bondi to show that
gravitational waves can have physical effects: in particular they suggested that if beads sliding on sticky rock move under the effect of the passing wave, they must transfer heat to the road by friction, which proves that gravitational waves carry energy (see e.g. Ref.\ \onlinecite{Berti:2016qkk}).}

The above examples suggest that our approach {provides} a simple interpretation of the interaction between GWs and test masses. Indeed, it is important to emphasise that current detectors are essentially looking for gravitoelectric effects. A proposal to exploit the gravitomagnetic effect is discussed in Ref. \  \onlinecite{Ruggiero_2020b}.

\section{Conclusions}\label{sec:conc}

Gravitational waves are today a key ingredient of multi-messenger astronomy, which allows us to study astrophysical sources using different channels and contributes to increasing our understanding of the Universe. As a consequence, we believe that it is important to teach the  main features of gravitational waves science in introductory physics and astronomy courses, which is useful also to give students the possibility of understanding the continuous breakthroughs in this field. The standard approach in teaching gravitational waves physics is usually based on  TT coordinates. On  one hand, these coordinates are useful because they  explain some basic characteristics of GWs, such as their polarizations and the fact that they are transverse to the propagation direction. On the other hand, TT coordinates lack a physical meaning and, in order to understand the interaction of the waves with detectors, it is necessary to obtain observable quantities, using the standard GR approach. 
From a teaching perspective, we believe that a different approach, based on the use of Fermi coordinates, would be more suitable: in fact, Fermi coordinates are  defined, by construction, as scalar invariants and have a concrete meaning,  since they are the coordinates an observer would naturally use to make space and time measurements in the vicinity of {her/his} world-line. Moreover, Fermi coordinates enable  simple understanding of the meaning of  the principle of equivalence, on which GR is based.  

We have shown that, thanks to Fermi coordinates, it is possible to describe the effects of a plane gravitational wave  using an electromagnetic analogy: in fact, the wave field is equivalent to the action of a gravitoelectric and a gravitomagnetic field, that are transverse to the propagation direction and orthogonal to each other. Moreover, the action of the wave on test masses is described in terms of  tidal forces, determined by a Lorentz-like equation. Then, it is easy to describe how the physical distance between two test masses changes, due to the passage of the wave: this can be understood as the action of a gravitoelectric field which, because of its tidal character, provokes  different effects on masses located at different positions. 
Furthermore, on the basis of this approach, it is possible to see that there are also gravitomagnetic effects, caused by the passage of the wave on moving test masses. Indeed, even if existing detectors, such as LIGO and VIRGO, and future ones, such as LISA, are  aimed at detecting gravitoelectric effects, it is possible that new types of detectors  could be designed to measure also gravitomagnetic effects. 

In summary, we believe that  our approach, which rests upon the analogy with well known facts from electromagnetic theory, could help students to better understand and explore gravitational waves physics.

\begin{acknowledgments}
{The author thanks Dr. Antonello Ortolan for the stimulating and useful discussions; moreover, the author expresses appreciation for the referees and the editors, whose suggestions greatly improved the paper.}
\end{acknowledgments}

\newpage

\begin{thebibliography}{100}

\bibitem{Einstein:1916cc}
A.~Einstein.
\newblock {Approximative Integration of the Field Equations of Gravitation}.
\newblock {\em Sitzungsber. Preuss. Akad. Wiss. Berlin (Math. Phys. )},
  1916:688--696, 1916.

\bibitem{Einstein:1918btx}
A.~Einstein.
\newblock {\"Uber Gravitationswellen}.
\newblock {\em Sitzungsber. Preuss. Akad. Wiss. Berlin (Math. Phys. )},
  1918:154--167, 1918.
  
\bibitem{Hulse:1974eb}
R.~Hulse and J.~Taylor.
\newblock {Discovery of a pulsar in a binary system}.
\newblock {\em Astrophys. J. Lett.}, 195:L51--L53, 1975.  

\bibitem{Weisberg:2004hi}
Joel~M. Weisberg and Joseph~H. Taylor.
\newblock {Relativistic binary pulsar B1913+16: Thirty years of observations
  and analysis}.
\newblock {\em ASP Conf. Ser.}, 328:25, 2005.

\bibitem{abbott2016observation}
Benjamin~P Abbott, Richard Abbott, TD~Abbott, MR~Abernathy, Fausto Acernese,
  Kendall Ackley, Carl Adams, Thomas Adams, Paolo Addesso, RX~Adhikari, et~al.
\newblock Observation of gravitational waves from a binary black hole merger.
\newblock {\em Physical review letters}, 116(6):061102, 2016.

\bibitem{burns2019opportunities}
E.~Burns, A.~Tohuvavohu, J.~M. Bellovary, E.~Blaufuss, T.~J. Brandt, S.~Buson,
  R.~Caputo, S.~B. Cenko, N.~Christensen, J.~W. Conklin, F.~D'Ammando, K.~E.~S.
  Ford, A.~Franckowiak, C.~Fryer, C.~M. Hui, K.~Holley-Bockelmann, T.~Jaffe,
  T.~Kupfer, M.~Karovska, B.~D. Metzger, J.~Racusin, B.~Rani, M.~Santander,
  J.~Tomsick, and C.~Wilson-Hodge.
\newblock Opportunities for multimessenger astronomy in the 2020s, 2019.

\bibitem{Farr:2011be}
Benjamin Farr, GionMatthias Schelbert, and Laura Trouille.
\newblock {Gravitational-wave Science in the High School Classroom}.
\newblock {\em arXiv.org}, September 2011.

\bibitem{d1992introducing}
R.~A. d'Inverno.
\newblock {\em Introducing Einstein's relativity}.
\newblock Clarendon Press, 1992.

\bibitem{schutz2009first}
Bernard Schutz.
\newblock {\em A first course in general relativity}.
\newblock Cambridge university press, 2009.

\bibitem{Centrella:2003gs}
Joan~M Centrella.
\newblock {Resource Letter: GrW-1: Gravitational waves}.
\newblock {\em American Journal of Physics}, 71(6):520--525, June 2003.

\bibitem{schutz1984gravitational}
Bernard~F Schutz.
\newblock Gravitational waves on the back of an envelope.
\newblock {\em American Journal of Physics}, 52(5):412--419, 1984.

\bibitem{Saulson:1997cg}
Peter~R Saulson.
\newblock {If light waves are stretched by gravitational waves, how can we use
  light as a ruler to detect gravitational waves?}
\newblock {\em American Journal of Physics}, 65(6):501--505, June 1997.

\bibitem{Melissinos:2010bv}
Adrian Melissinos and Ashok Das.
\newblock {The response of laser interferometers to a gravitational wave}.
\newblock {\em American Journal of Physics}, 78(11):1160--1164, November 2010.

\bibitem{Mathur:2017cs}
Harsh Mathur, Katherine Brown, and Ashton Lowenstein.
\newblock {An analysis of the LIGO discovery based on introductory physics}.
\newblock {\em American Journal of Physics}, 85(9):676--682, September 2017.

\bibitem{Buskirk:2019ho}
Dillon Buskirk and Maria C~Babiuc Hamilton.
\newblock {A complete analytic gravitational wave model for undergraduates}.
\newblock {\em EUROPEAN JOURNAL OF PHYSICS}, 40(2):025603--20, January 2019.

\bibitem{Burko:2017ex}
Lior~M Burko.
\newblock {Gravitational Wave Detection in the Introductory Lab}.
\newblock {\em The Physics Teacher}, 55(5):288--292, May 2017.

\bibitem{Garfinkle:2006bz}
David Garfinkle.
\newblock {Gauge invariance and the detection of gravitational radiation}.
\newblock {\em American Journal of Physics}, 74(3):196--199, March 2006.

\bibitem{de2010classical}
Fernando De~Felice and Donato Bini.
\newblock {\em Classical measurements in curved space-times}.
\newblock Cambridge University Press, 2010.

\bibitem{Rakhmanov_2014}
Malik Rakhmanov.
\newblock Fermi-normal, optical, and wave-synchronous coordinates for spacetime
  with a plane gravitational wave.
\newblock {\em Classical and Quantum Gravity}, 31(8):085006, apr 2014.

\bibitem{flanagan2005basics}
Eanna~E Flanagan and Scott~A Hughes.
\newblock The basics of gravitational wave theory.
\newblock {\em New Journal of Physics}, 7(1):204, 2005.

\bibitem{1922RendL..31...21F}
E.~{Fermi}.
\newblock {Sopra i fenomeni che avvengono in vicinanza di una linea oraria}.
\newblock {\em Rend.~Lincei, 1922, 31(1), pp.~21-23, 51-52, 101-103 ( in
  Italian)}, 31:21--23, December 1922.
  
  \bibitem{synge1960relativity}
J.L. Synge.
\newblock {\em Relativity: The General Theory}.
\newblock Number v. 1 in North-Holland series in physics. North-Holland
  Publishing Company, 1960.
  
\bibitem{Ruggiero_2020}
Matteo~Luca Ruggiero and Antonello Ortolan.
\newblock Gravito-electromagnetic approach for the space-time of a plane
  gravitational wave.
\newblock {\em Journal of Physics Communications}, 4(5):055013, may 2020.  

\bibitem{McDonald:1997fd}
Kirk~T McDonald.
\newblock {Answer to Question {\#}49. Why c for gravitational waves?}
\newblock {\em American Journal of Physics}, 65(7):591--592, July 1997.  

\bibitem{Ruggiero:2002hz}
Matteo~Luca Ruggiero and Angelo Tartaglia.
\newblock {Gravitomagnetic effects}.
\newblock {\em Nuovo Cim.}, B117:743--768, 2002.

\bibitem{Mashhoon:2003ax}
Bahram Mashhoon.
\newblock {Gravitoelectromagnetism: A Brief review}.
\newblock 2003.

\bibitem{biniortolan2017}
Donato Bini, Andrea Geralico, and Antonello Ortolan.
\newblock Deviation and precession effects in the field of a weak gravitational
  wave.
\newblock {\em Phys. Rev. D}, 95:104044, May 2017.

\bibitem{Ruggiero_2020b}
Matteo~Luca Ruggiero and Antonello Ortolan.
\newblock Gravitomagnetic resonance in the field of a gravitational wave.
\newblock {\em Phys. Rev. D}, 102:101501, Nov 2020.

\bibitem{duit1991role}
Reinders Duit.
\newblock On the role of analogies and metaphors in learning science.
\newblock {\em Science education}, 75(6):649--672, 1991.

\bibitem{venville1996role}
Grady~J Venville and David~F Treagust.
\newblock The role of analogies in promoting conceptual change in biology.
\newblock {\em Instructional Science}, 24(4):295--320, 1996.

\bibitem{MTW}
Charles~W Misner, Kip~S Thorne, and John~A Wheeler.
\newblock {\em Gravitation}.
\newblock San Francisco: WH Freeman and Co., 1973.

\bibitem{Mashhoon:1996wa}
Bahram Mashhoon, James~C. McClune, and Hernando Quevedo.
\newblock {Gravitational superenergy tensor}.
\newblock {\em Phys. Lett. A}, 231:47--51, 1997.

\bibitem{marzlin}
Karl-Peter Marzlin.
\newblock Fermi coordinates for weak gravitational fields.
\newblock {\em Phys. Rev. D}, 50:888--891, Jul 1994.

\bibitem{manasse1963fermi}
FK~Manasse and Charles~W Misner.
\newblock Fermi normal coordinates and some basic concepts in differential
  geometry.
\newblock {\em Journal of mathematical physics}, 4(6):735--745, 1963.

\bibitem{press1972gravitational}
William~H Press and Kip~S Thorne.
\newblock Gravitational-wave astronomy.
\newblock {\em Annual Review of Astronomy and Astrophysics}, 10(1):335--374,
  1972.

\bibitem{braginskij1969reception}
VB~Braginskij, Ya~B Zel'Dovich, and VN~Rudenko.
\newblock Reception of gravitational radiation of extraterrestrial origin.
\newblock {\em Soviet Journal of Experimental and Theoretical Physics Letters},
  10:280--283, 1969.





\bibitem{braginsky1971heterodyne}
R.~W. Davies, editor.
\newblock {\em On the heterodyne method of detecting gravitational waves},
  volume Proceedings of the Conference on experimental tests of gravitation
  theories, 1971.  
  
\bibitem{Berti:2016qkk}
Emanuele Berti.
\newblock {The First Sounds of Merging Black Holes}.
\newblock {\em APS Physics}, 9:17, 2016.  

\end{thebibliography}
\end{document}